\documentclass[aps,prd,twocolumn,tightenlines,preprintnumbers,showpacs,superscriptaddress,notitlepage,multicol,nofootinbib,eqsecnum,floatfix,longbibliography,10pt]{revtex4-1}

\usepackage{amsmath, amssymb}
\usepackage{cancel}
\usepackage{graphicx}  
\usepackage{dcolumn}   
\usepackage{bm}        
\usepackage{standalone}
\usepackage{enumitem}
\usepackage[pdftex]{color}
\usepackage[utf8]{inputenc}
\usepackage{xcolor}
\usepackage{xspace}
\usepackage{hyperref}
\hypersetup{
    colorlinks=true,  
    linkcolor=blue,   
    citecolor=blue,   
    filecolor=blue,   
    urlcolor=blue     
}

\usepackage{placeins}

\usepackage{slashed}

\newcommand\pslash{\slashed{p}}

\newcommand\kslash{\slashed{k}}
\def\ptwoslash{\slashed{p}_2}
\def\pthreeslash{\slashed{p}_3}
\newcommand\Tr{\mathop{\text{Tr}}}
\newcommand\eps{\epsilon}
\def\epsbar{\overline{\epsilon}}

\def\afpC{\frac\alpha{4\pi} C_F}
\def\mbar{\ensuremath{{\overline{m}}}\xspace}
\def\msbar{\ensuremath{{\overline{\text{MS}}}}\xspace}

\def\Zqmsmom{Z^\text{mSMOM}_q}

\def\ZPmsmom{Z^\text{mSMOM}_\text{P}}
\def\Zmmsmom{Z^\text{mSMOM}_m}

\newcommand{\gS}{\mathrm{S}}
\newcommand{\gP}{\mathrm{P}}
\newcommand{\gV}{\mathrm{V}}
\newcommand{\gA}{\mathrm{A}}

\newcommand{\smom}{\ensuremath{\mathrm{SMOM}}\xspace}
\newcommand{\msmom}{\ensuremath{\mathrm{mSMOM}}\xspace}

\newcommand{\ms}{\ensuremath{\overline{\mathrm{MS}}}\xspace}
\newcommand{\avg}[1]{\left< #1 \right>} 

\newcommand\cern{CERN, Theoretical Physics Department, Geneva, Switzerland}
\newcommand\higgs{Higgs Centre for Theoretical Physics, School of
  Physics \& Astronomy, The University of Edinburgh, Edinburgh EH9 3FD, UK}
\newcommand\soton{School of Physics and Astronomy, University of
  Southampton,  Southampton SO17 1BJ, UK}

\begin{document}
\title{Absorbing discretisation effects with a massive renormalization scheme: the charm-quark mass}

\author{L.~Del~Debbio}\affiliation{\higgs}
\author{F.~Erben}\affiliation{\cern}
\author{J.~M.~Flynn}\affiliation{\soton}
\author{R.~Mukherjee}\affiliation{\soton}
\author{J.~T.~Tsang}\thanks{Corresponding author}\affiliation{\cern}

\collaboration{RBC and UKQCD Collaborations}\noaffiliation
\begin{abstract}
  We present the first numerical implementation of the massive SMOM (\msmom)
  renormalization scheme and use it to calculate the charm quark mass. Based on
  ensembles with three flavours of dynamical domain wall fermions with lattice
  spacings in the range $0.11$ -- $0.08\,\mathrm{fm}$, we demonstrate that the
  mass scale which defines the \msmom scheme can be chosen such that the
  extrapolation has significantly smaller discretisation effects than the \smom
  scheme. Converting our results to the $\overline{\mathrm{MS}}$ scheme we
  obtain $\overline{m}_c(3\,\mathrm{GeV}) = 1.008(13)\,\mathrm{GeV}$ and
  $\overline{m}_c(\overline{m}_c) = 1.292(12)\,\mathrm{GeV}$.
\end{abstract}

\maketitle
\preprint{CERN-TH-2024-124}

\makeatletter
\def\l@subsection#1#2{}
\def\l@subsubsection#1#2{}
\makeatother

\tableofcontents
\section{Introduction} 
Non-perturbative massive renormalization schemes, such as the ones introduced in
Ref.~\cite{Boyle:2016wis}, yield renormalized correlators that satisfy vector
and axial Ward identities independently of the value of the quark masses and are
expected to reabsorb some of the lattice artefacts that come in powers of $am$
and can be large for heavy quark masses. These schemes are therefore interesting
candidates to renormalize quantities that are affected by large cut-off effects,
leading to milder extrapolations to the continuum limit compared to the usual
massless schemes that are currently used.

In this paper, we present the first numerical implementation of the
renormalization conditions that were spelled out in Ref.~\cite{Boyle:2016wis}
and extract the renormalization constants that are needed in order to compute
the renormalized charm-quark mass in these massive
symmetric-momentum-subtraction (mSMOM) schemes. The schemes are labelled by the
momentum scale of the subtraction point and by the value of the renormalized
quark mass at which the renormalization conditions are imposed. Lattice
artefacts depend on the choice of this mass, which can be tuned in order to
obtain flatter extrapolations. We use lattice QCD ensembles generated by the
RBC/UKQCD collaboration, with 2+1 dynamical flavours and inverse lattice spacings
ranging from $a^{-1}=1.73~\mathrm{GeV}$ to $2.79~\mathrm{GeV}$. We compute all
the lattice correlators that enter the renormalization conditions and spell out
in detail the workflow to implement and solve the correct set of equations.

Results in different mSMOM schemes are converted to \msbar using one-loop
conversion factors and show a pleasing consistency. The main result of this
first study confirms the theoretical expectation motivating massive
schemes. They provide a (simple) way to absorb some of the mass-dependent
lattice artefacts and yield more reliable extrapolations to the continuum limit.

The remainder of this paper is organised as follows. In Section~\ref{sec:mnpr}
we remind the reader of the details of the massive non-perturbative
renormalization scheme. In Section~\ref{sec:ens} we provide details of our
numerical set-up before presenting the details of our analysis and our final
results in Section~\ref{sec:anal}. We conclude with an outlook in
section~\ref{sec:conc}. An early stage of this analysis was reported in
Ref.~\cite{DelDebbio:2023naa}

\section{Massive NPR}
\label{sec:mnpr}
Before discussing the numerical analysis that was performed for this paper, we
summarise the main ideas behind massive renormalization schemes. To keep our
presentation self-contained, we quote below the renormalization conditions
defining \msmom schemes, which were originally spelled out in
Ref.~\cite{Boyle:2016wis}. To match the numerical simulations, we work in
Euclidean space. In our conventions, bare quantities are written without any
suffix, while their renormalized counterparts are identified by a suffix
$R$. The renormalization conditions are usually expressed in terms of amputated
correlators of fermion bilinears
\begin{equation}
  \label{eq:LambdaO}
  \Lambda^a_\Gamma(p_2,p_3) = S(p_3)^{-1} G^a_\Gamma(p_3,p_2) S(p_2)^{-1}\, ,
\end{equation}
where $S(p)$ is the fermion propagator,
\begin{equation}
  \label{eq:Sdef}
  S(p) = \int d^4x\, e^{-ip\cdot x} \langle \psi(x)\bar\psi(0)\rangle\, ,
\end{equation}
and $G^a_\Gamma(p_3,p_2)=\langle O^a_\Gamma(q)\bar\psi(p_3)\psi(p_2) \rangle$
with $O^a_\Gamma = \bar\psi \Gamma\tau^a\psi$.  The superscript $a$, which will
be dropped henceforth, denotes that we consider flavour non-singlet bilinears,
with $\tau^a$ a generic flavour-rotation generator. In the following we will
also suppress the superscript \msmom unless it is required to avoid ambiguity.

We choose the same symmetric momentum configurations as those chosen in in the
massless \smom scheme, i.e.  $q^2\equiv (p_2-p_3)^2=p_2^2=p_3^3=\mu^2$, where
$\mu$ is the renormalization scale. The massive scheme requires the introduction
of another scale, $\overline{m}_R$, a renormalized mass at which the
renormalization conditions are imposed. The massless scheme is recovered in the
limit $\overline{m}_R\to 0$. For the \msmom scheme in Euclidean space the
renormalization conditions, to be evaluated with the symmetric momentum
configuration imposed and at $m_R=\overline{m}_R$, read
\begin{align}
  1&=\frac{1}{12 p^2} \operatorname{Tr}\left[-i S_R(p)^{-1} \cancel{p}\right], \label{eq:Z_q}\\
  1&=\frac{1}{12 m_R}\Big\{\operatorname{Tr}\left[S_R(p)^{-1}\right]+\frac{1}{2} \operatorname{Tr}\left[\left(iq \cdot \Lambda_{\mathrm{A}, R}\right) \gamma_5\right]\Big\}, \label{eq:Z_m}\\
  1&=\frac{1}{12 q^2} \operatorname{Tr}\left[\left(q \cdot \Lambda_{\mathrm{V}, R}\right) \cancel{q}\right], \label{eq:Z_V}\\
  1&=\frac{1}{12 q^2} \operatorname{Tr}\left[\left(q \cdot \Lambda_{\mathrm{A}, R}\, +\, 2m_R\Lambda_{\mathrm{P},R}\right) \gamma_5 \cancel{q}\right], \label{eq:Z_A}\\
  1&=\frac{1}{12i} \operatorname{Tr}\left[\Lambda_{\mathrm{P}, R} \gamma_5\right], \label{eq:Z_P}\\
  1&=\frac{1}{12} \operatorname{Tr}\left[\Lambda_{\mathrm{S}, R}\right] +\, \frac{1}{6q^2}\operatorname{Tr}\left[2m_R\Lambda_{\mathrm{P},R}\gamma_5\cancel{q}\right]. \label{eq:Z_S}
\end{align}

The renormalized quantities are defined as follows: 
\begin{gather}
  \label{eq:Zdef}
  \psi_R = Z_q^{1/2} \psi\, , \quad m_{R} = Z_{m} m\, , \quad O_{\Gamma,R} = Z_\Gamma O_\Gamma\, ,
\end{gather}
where $m$ denotes a quark mass. The renormalized propagator and amputated vertex
functions are 
\begin{equation}
  \label{eq:SGren}
  S_R(p) = Z_q S(p)\, , \quad \Lambda_{\Gamma,R}(p_2,p_3) =
  \frac{Z_\Gamma}{Z_q} \Lambda_\Gamma(p_2,p_3)\, .
\end{equation}
As discussed in the original publication~\cite{Boyle:2016wis}, these conditions
ensure that renormalized correlators satisfy the Ward identities of the
continuum theory, which in turn lead to useful constraints on the
renormalization constants\footnote{In~\cite{Boyle:2016wis} it was checked that
  the last condition in Eq.~\eqref{eq:Z_S} ensured $Z_\gS=Z_\gP$ at one loop in
  continuum perturbation theory in Feynman gauge. For other gauge choices, this
  condition should be modified. This renormalization condition is not used in
  the analysis presented in this paper.}, namely
\begin{equation}
  \label{eq:Zrelations}
  Z_\gV = Z_\gA =1, \quad Z_\gP=Z_\gS, \quad Z_m Z_\gP=1\, .
\end{equation}

Substituting Eqs.~\eqref{eq:Zdef} and~\eqref{eq:SGren} into the renormalization
conditions Eqs.~\eqref{eq:Z_q}--\eqref{eq:Z_S} and solving the system of
equations gives access to the renormalization factors $Z_q$, $Z_m$, $Z_A$,
$Z_V$, $Z_S$ and $Z_P$. In practice we find it convenient to replace the
renormalization condition Eq.~\eqref{eq:Z_q} by a direct determination of $Z_A$
from ratios of conserved and local axial currents. Combined with
Eqs.~\eqref{eq:Z_m}--\eqref{eq:Z_S} this still gives access to all the required
renormalization constants.

Note that by construction the renormalization constants in a massive scheme
depend on both the coupling and the dimensionless product $a \overline{m}$. The
\msmom schemes are defined by tuning the renormalized quark mass to some
arbitrary scale $\overline{m}_R$, where the renormalization conditions need to
be satisfied.

The arbitrariness in the choice of $\overline{m}_R$ can be turned into a useful
tool when extrapolating lattice QCD results to the continuum limit. Indeed, the
\emph{ideal} choice of $\overline{m}_R$ is determined by requiring that the
observables of interest have a mild dependence on the lattice spacing in that
particular scheme. Different observables may dictate different values of
$\overline{m}_R$; this is not a problem, since we know how to connect schemes
corresponding to different choices of $\overline{m}_R$ to a common reference
scheme such as, e.g., \ms, using the one-loop perturbative expressions in
Ref.~\cite{Boyle:2016wis} and in Appendix~\ref{app:mass-conversion}.

The focus of this paper is to compute the renormalized charm-quark mass in
\msmom, which is defined as
\begin{equation}
  \label{eq:RenormMassDef}
  m^\msmom_{c,R}(\mu,\overline{m}_R) = \lim_{a\to 0}Z_m^\msmom(g,a\mu,a\overline{m}) (am_c) a^{-1}\,,\\
\end{equation}
where the mass scale $\overline{m}_R$ defining the renormalization scheme is
obtained through
\begin{equation}
  \label{eq:RenormMassScaleDef}
  \overline{m}_R(\mu,\overline{m}_R) = \lim_{a\to 0}Z_m^\msmom(g,a\mu,a\overline{m}) (a\overline{m}) a^{-1}\,,
\end{equation}
and the bare quark mass in lattice units ($am$) is the sum of the input quark
mass $am_q$ and the additive mass renormalization $am_\mathrm{res}$
\begin{equation}
  am  \equiv \left(am_q + am_\mathrm{res}\right)\,,
\end{equation}
$Z_m^\msmom(g,a\mu,a\overline{m}_R)$ is the renormalization constant defined by
the renormalization conditions above and the bare mass of the charm quark is set
by requiring that the mass of the heavy-heavy pseudoscalar meson coincides with
the mass of the physical $\eta_c$ meson.

After taking the continuum limit, the mSMOM renormalized mass can be converted
to \msbar,
\begin{equation}
m_R^\msbar(\tilde\mu) = R^{\msbar\leftarrow\text{mSMOM}}(\tilde\mu,\mu,\mbar_R) m_R^\text{mSMOM}(\mu,\mbar_R),
\end{equation}
where the dimensionful scale $\tilde\mu$ stems from dimensional regularisation
and will in practice be set equal to $\mu$.  The conversion factor
$R^{\msbar\leftarrow\text{mSMOM}} = Z_m^\msbar/\Zmmsmom$ is evaluated at
one-loop in perturbation theory in Appendix~\ref{app:mass-conversion}.

\section{Simulation set up and strategy \label{sec:ens}}

\begin{table}
  \caption{Summary of the main parameters of the ensembles used in this work.
    In the ensemble name the first letter (C, M or F) stands for coarse, medium
    and fine, respectively. The last letter (M or S) stands for M\"obius and
    Shamir kernels, respectively.}
  \label{tab:enspar}
  \begin{tabular}{lcrccll}
	 \hline
	 \hline
	 name & $L/a$ & $T/a$ & $a^{-1}[\textrm{GeV}]$ & $M_\pi[\textrm{MeV}]$ & $am_l$  & $am_s$\\
	 \hline
	 C1M & $24$ & $64$ & 1.7295(38) & 276 & 0.005   & 0.0362 \\
	 C1S & $24$ & $64$ & 1.7848(50) & 340 & 0.005   & 0.04   \\
	 \hline
	 M0M & $64$ & $128$ & 2.3586(70) & 139 & 0.000678  & 0.02661 \\
	 M1M & $32$ & $64$ & 2.3586(70) & 286 & 0.004   & 0.02661 \\
	 M1S & $32$ & $64$ & 2.3833(86) & 304 & 0.004   & 0.03    \\
	 \hline
	 F1M & $48$& $96$ & 2.708(10)  & 232 & 0.002144 & 0.02144 \\
	 F1S & $48$& $96$ & 2.785(11)  & 267 & 0.002144 & 0.02144 \\
	 \hline
	 \hline
  \end{tabular}
\end{table}

\begin{table}
  \caption{Heavier input quark masses that were simulated in addition to $am_l$, $2am_l$, $am_s/2$ and $am_s$.}
  \label{tab:ensmasses}
  \resizebox{\columnwidth}{!}{
    \begin{tabular}{cc}
      \hline
      \hline
      ens & $am_q$ \\
      \hline
      C1M  & 0.05, 0.1, 0.15, 0.2, 0.3\\
      C1S  &0.05, 0.1, 0.15, 0.2, 0.3, 0.33\\
      \hline
      M1M  & 0.05, 0.1, 0.15, 0.225, 0.3, 0.32, 0.34\\
      M1S  & 0.05, 0.1, 0.15, 0.225, 0.3, 0.32, 0.34, 0.36, 0.375\\
      \hline
      F1M  & 0.033, 0.066, 0.099, 0.132, 0.198, 0.264, 0.33, 0.36\\
      F1S  & 0.033, 0.066, 0.099, 0.132, 0.198, 0.264, 0.33, 0.36, 0.396\\
      \hline
      \hline
    \end{tabular}
  }
\end{table}

We use RBC/UKQCD's ensembles~\cite{Aoki:2010pe, RBC:2014ntl, Boyle:2017jwu,
  Boyle:2018knm, Boyle:2024gge} with Iwasaki gauge
action~\cite{Iwasaki:1984cj,Iwasaki:1985we} and domain-wall fermion
action~\cite{Shamir:1993zy,Furman:1994ky}. They include the dynamical effects
from degenerate up and down quarks as well as the strange quark. The main
ensemble properties are listed in Table~\ref{tab:enspar}. For each of the three
lattice spacings we have one ensemble with the Shamir domain-wall
kernel~\cite{Shamir:1993zy} (last letter `S') and one with the M\"obius
domain-wall kernel~\cite{Brower:2004xi, Brower:2005qw, Brower:2012vk} (last
letter `M'). The parameters of these kernels are chosen such that a combined
continuum limit with all ensembles is possible~\cite{RBC:2014ntl}. In addition
we have data around the physical charm quark mass on the physical pion mass
ensemble M0M which differs from M1M only in pion mass and volume.

We implement the SMOM momentum configuration by choosing momenta $p_2 =
(p,p,0,0)$ and $p_3=(p,0,p,0)$ where $p = \frac{2\pi}{L} (n+\theta)$. Since our
aim is to comprehensively cover the region $2\,\mathrm{GeV} \lesssim q \lesssim
3\,\mathrm{GeV}$ we use twist angles $\theta \in \{0,0.25,0.5,0.75\}$ in
combination with Fourier modes $n\in\{3,4,5\}$ for the coarse and medium, and
$n\in\{4,5,6\}$ for the fine ensembles.

We map out the parameter space by simulating at several quark masses $am_q$
between the light-quark mass and the largest quark mass we can reach on a given
ensemble whilst maintaining good control over the residual-mass determination
and the domain-wall formalism~\cite{Boyle:2016imm}. Since we expect sea-pion
mass and finite volume effects to be negligible for the determination of the
charm quark mass, the main numerical analysis is based on the computationally
cheaper non-physical pion mass ensembles. However, in order to assess these
effects, we simulated a small number of heavy quark masses in the charm region
on the M0M ensemble which can be directly compared with the equivalent M1M
datapoints. As we will see in Sec.~\ref{subsec:obs} sea-pion mass effects are at
the sub-permille level.

The chosen quark masses are listed in Table~\ref{tab:ensmasses}. The
measurements were carried out using the Grid and Hadrons
libraries~\cite{Boyle:2016lbp, Yamaguchi:2022feu,
  antonin_portelli_2023_8023716}.

For each input quark mass $am_q$ we compute vertex functions (see
Eq.~\eqref{eq:LambdaO}) as well as several mesonic flavour-diagonal
quark-connected two-point correlation functions. For the latter we use a mild
Jacobi smearing to improve the overlap with the ground state for heavy masses,
in particular for the pseudoscalar density $P$, the midpoint pseudoscalar
density $J_{5q}$~\cite{Furman:1994ky} and the local ($L$) and conserved
($C$)~\cite{Blum:2000kn,Boyle:2014hxa,RBC:2014ntl} versions of the temporal
component of the axial current. We determine the residual mass $am_\mathrm{res}$
and the renormalization constant $Z_A$ from the late time behaviour of ratios of
these correlation functions via
\begin{equation}
  am_\mathrm{res}^\mathrm{eff}(t) = \frac{\avg{PJ_{5q}}(t)}{\avg{PP}(t)}\,,
  \label{eq:amres}
\end{equation}
and
\begin{equation}
  Z_A^\mathrm{eff}(t) = \frac{1}{2} \left[\frac{C(t+\frac{1}{2})+C(t-\frac{1}{2})}{2L(t)} + \frac{2C(t+\frac{1}{2})}{L(t)+L(t+1)}\right]\,.
  \label{eq:ZAeff}
\end{equation}

Throughout this work we set the quark mass using the quark-connected
flavour-diagonal pseudoscalar meson $M_{\eta_h}$, since the quantity we are
ultimately interested in is the charm quark mass and the contribution from
quark-disconnected pieces to the mass of the $\eta_c$ meson has been estimated
to be negligibly small~\cite{Davies:2010ip}. We explore reference masses in the
range $\frac{1}{2}M^\mathrm{PDG}_{\eta_c}$ to $M^\mathrm{PDG}_{\eta_c} =
2.9839(4)\,\mathrm{GeV}$~\cite{ParticleDataGroup:2022pth}.

The strategy of our calculation is as follows:
\begin{enumerate}[label=(\alph*)]
\item For each mass $am_q$ on each ensemble, determine $am_\mathrm{res}$ and
  hence $am$ as well as $aM_{\eta_h}(am)$, $Z_A(g,am)$, $Z_m(g,a\mu,am)$.
\item Interpolate $Z_m(g,a\mu,am)$ to a common momentum scale $\hat{\mu}$ to
  obtain $Z_m(g,a\hat{\mu},am)$ on all ensembles.\label{step1}
\item Fix two mass-scales: the scale $\overline{m}$ at which the renormalization
  conditions are imposed and the quark mass $m$ to be determined. These do not
  have to be the same.
  
  In practice we define a set of meson masses $M_i$ such that
  $M_i/M_{\eta_c}^\mathrm{PDG} \in \{0.5,0.6,0.7, 0.75,0.8,0.9,1\}$. We
  interpolate $am(aM)$ to each choice of $M_i$ to obtain $am_i$ and similarly
  interpolate $Z_m(g,a\hat{\mu},am)$ to obtain $Z_m(g,a\hat{\mu},am_i)$. We note
  that the heaviest two and three values of $M_i$ are not directly accessible on
  the C1S and C1M ensembles, respectively.\label{step2}

  Then, we define the mass-scale $\overline{m}$ of the renormalization
  condition by fixing a meson mass $\overline{M}$ to be one of the
  $M_i$ and set the bare quark mass $m$ by fixing a meson mass $M$ to
  a potentially different $M_i$.
\item For the given choice of $M$ and $\overline{M}$, combine
  $Z_m(g,a\hat{\mu},a\overline{m})$ and $am$ to obtain the right hand side of
  Eq.~\eqref{eq:RenormMassDef} on each ensemble. Take the continuum limit to
  obtain $m_R(\hat{\mu},\overline{m}_R)$. Finally, (since $M$ and $\overline{M}$
  can differ) also take the continuum limit to obtain
  $\overline{m}_R(\hat{\mu},\overline{m}_R)$
  (c.f. Eq.~\eqref{eq:RenormMassScaleDef}). This last step is required in order
  to know the mass scale of the renormalization condition which is needed to
  relate it to other schemes such as \smom or \ms.
  \label{step3}
\item Our choice of domain wall parameters does not allow for direct simulations
  at the physical charm quark mass on the coarse lattice spacing. Hence we
  repeat this procedure for different values of $M$, but at fixed
  $\overline{M}$. This yields values $m_{i,R}^\msmom(\hat{\mu},\overline{m}_R)$
  as a function of $M_i$, which can be parameterised to finally obtain the value
  of $m_{c,R}^\msmom(\hat{\mu},\overline{m}_R)$. \label{step4}
\item Finally, repeat the entire analysis for different choices of $\hat{\mu}$
  and $\overline{M}$ in order to determine the \emph{ideal} choice of
  $\overline{m}$ for a given $\hat{\mu}$.\label{step5}
\end{enumerate}

\section{Results \label{sec:anal}}
In this section we carry out the analysis outlined above.
\subsection{From correlators to observables \label{subsec:obs}}

\begin{figure}
  \includegraphics[width=\columnwidth]{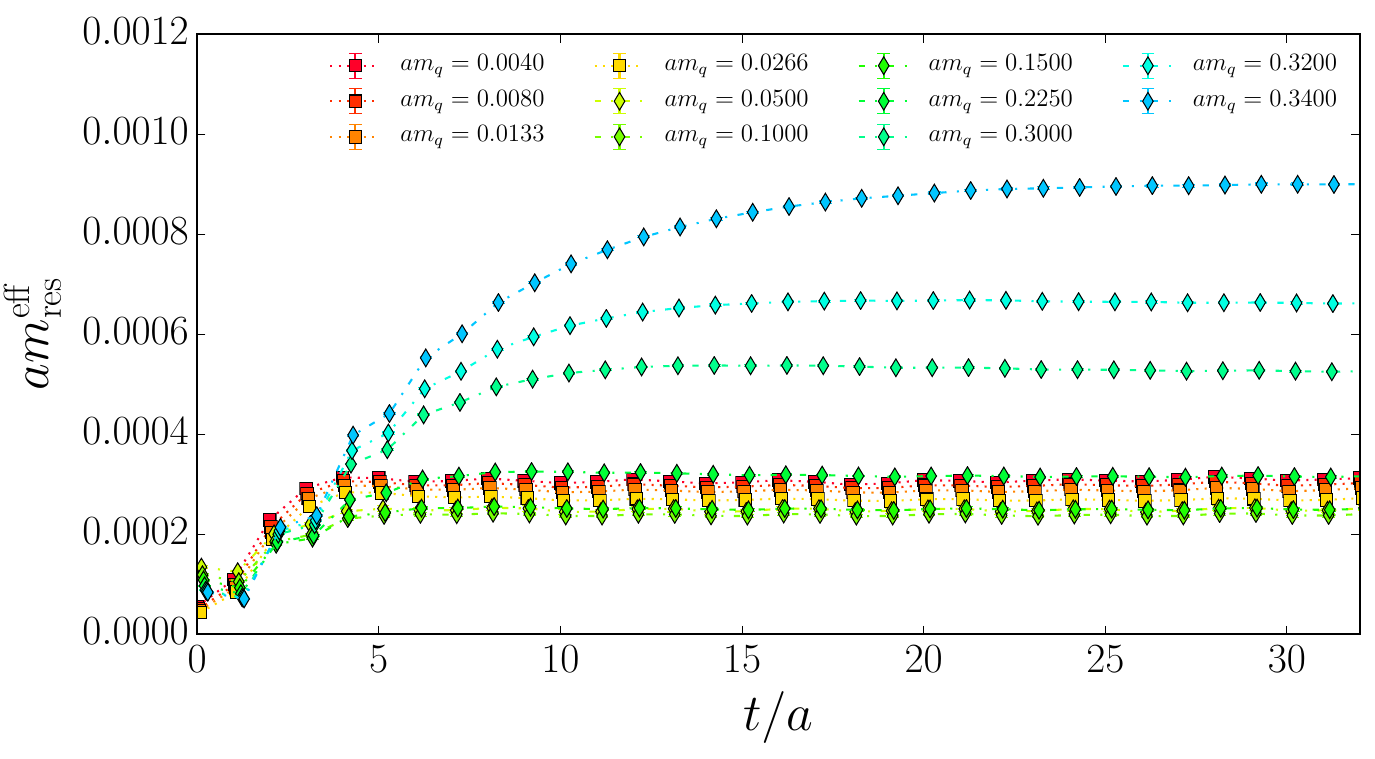}
  \includegraphics[width=\columnwidth]{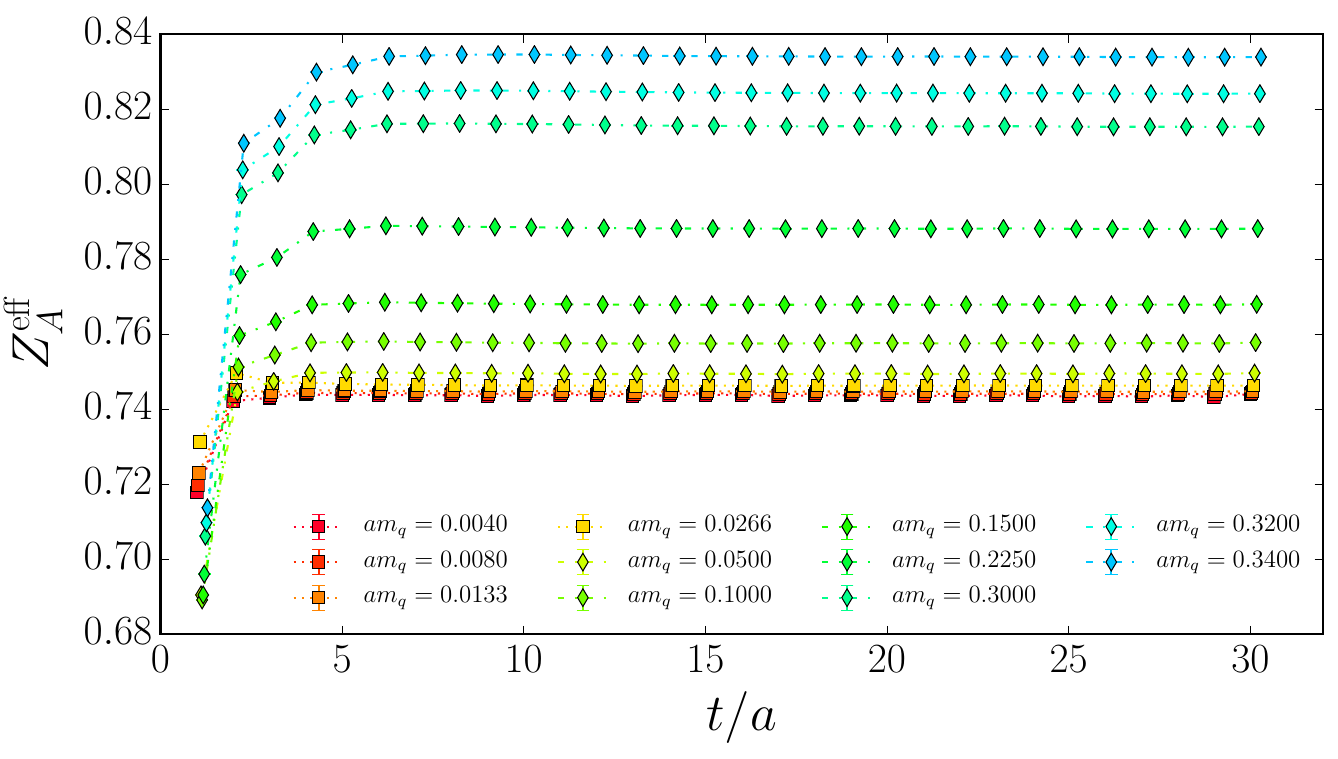}
  \includegraphics[width=\columnwidth]{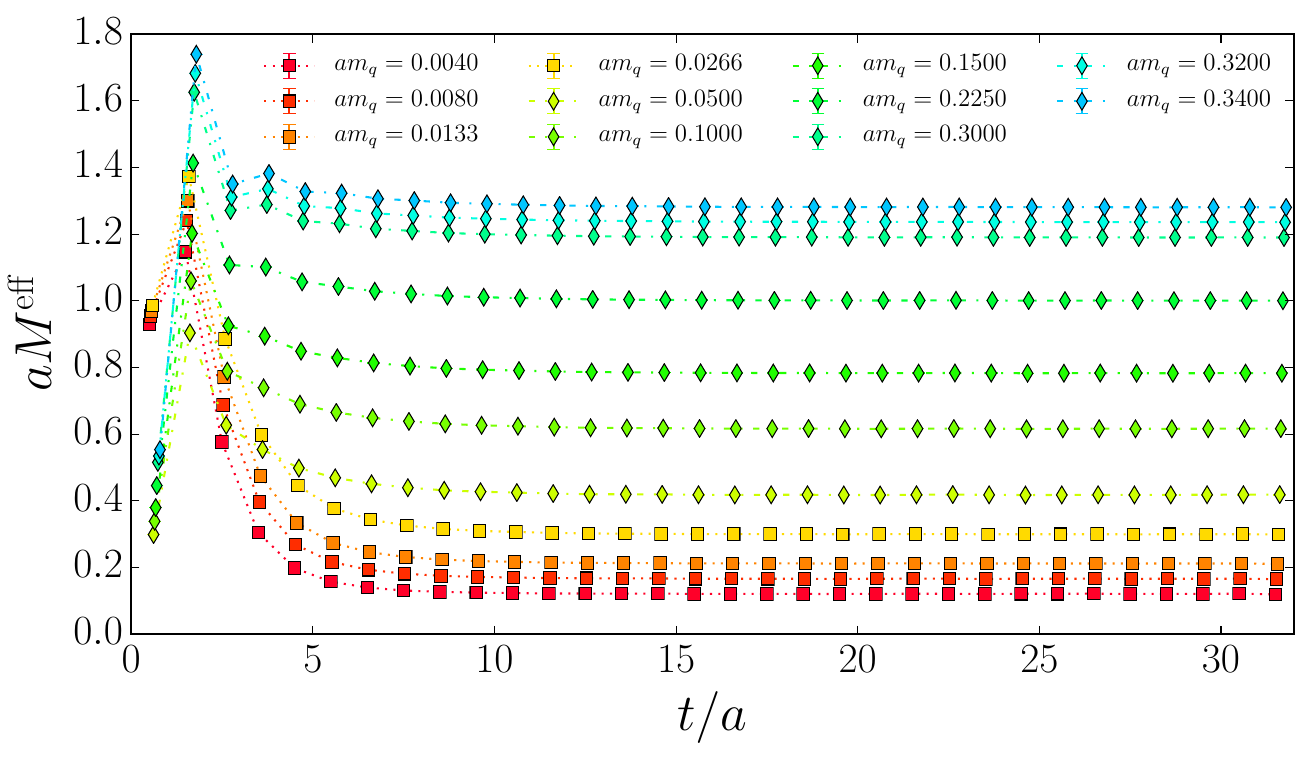}
  \caption{Representative effective $am_\mathrm{res}$ (top) and $Z_A$ (middle)
    and $aM_{\eta_h}$ (bottom)values on the M1M ensemble.}
  \label{fig:ZA_amres_M1M}
\end{figure}

\begin{figure}
  \includegraphics[width=\columnwidth]{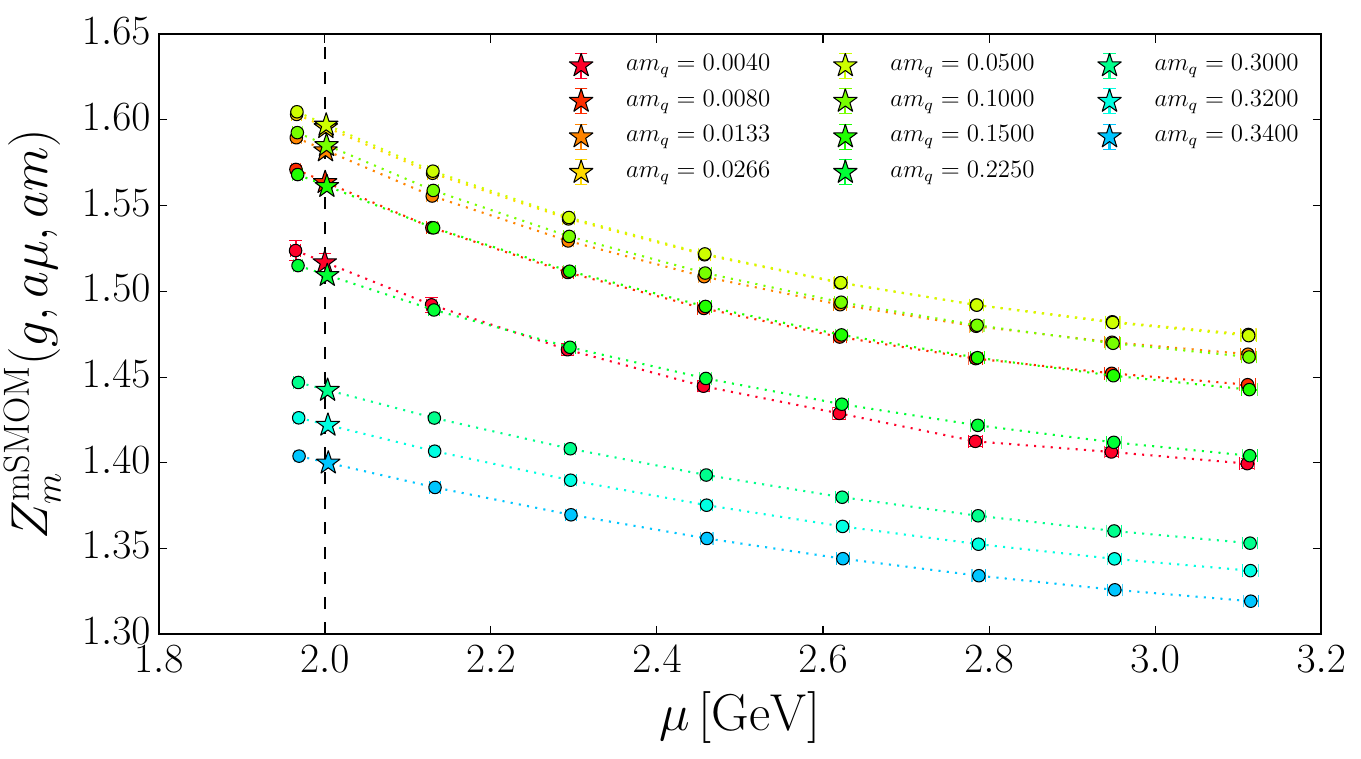}
  \caption{Interpolation of $Z_m$ to a scale of $\mu=2\,\mathrm{GeV}$ for
    various input quark mass $am_q$. The stars indicate the value of
    $Z_m^\msmom$ interpolated to $\hat{\mu}=2\,\mathrm{GeV}$.}
  \label{fig:Zminterp}
\end{figure}

\begin{figure}
  \includegraphics[width=\columnwidth]{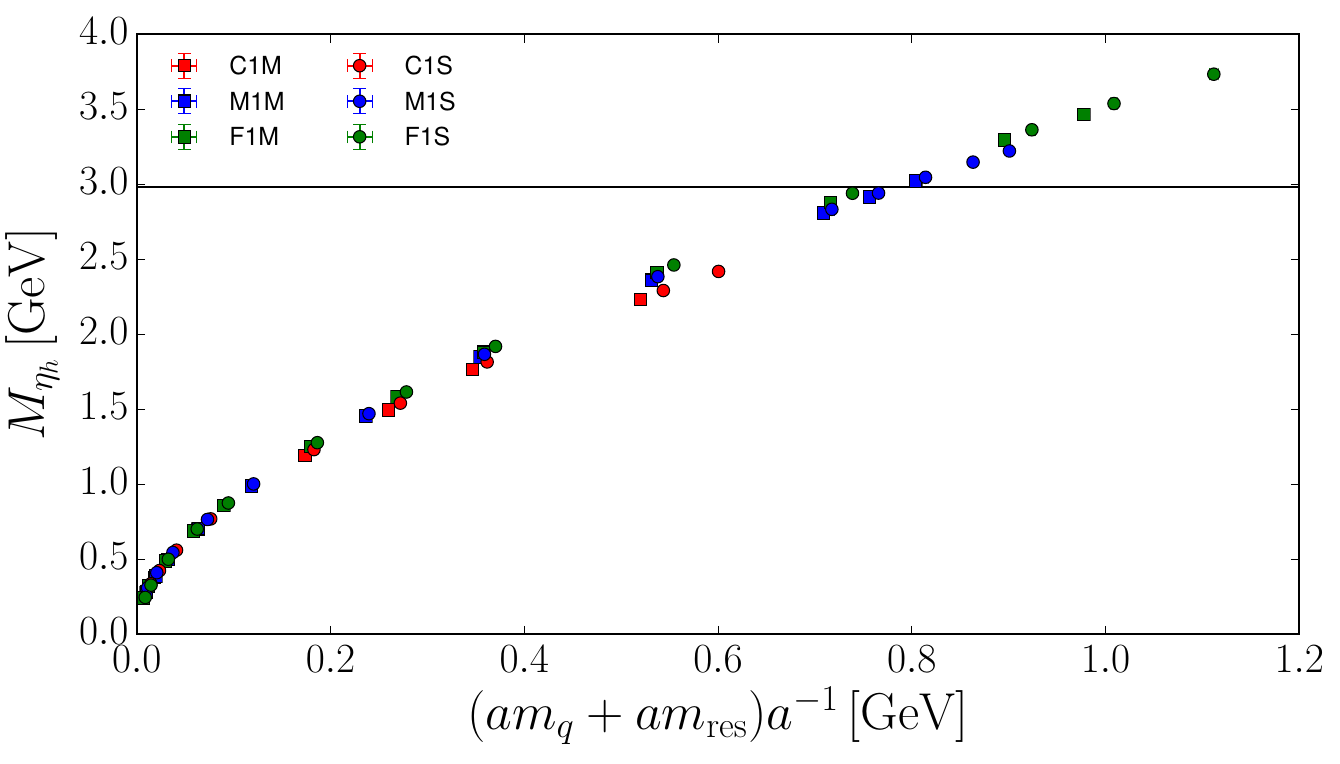}
  \caption{Coverage of the quark mass dependence of our data.}
  \label{fig:allmasses}
\end{figure}

We first determine $am_\mathrm{res}$, $Z_A$ and $aM$ on all ensembles and for
each choice of quark mass.  The plots in Figure~\ref{fig:ZA_amres_M1M}
illustrate the time behaviour of the data on the M1M ensemble
(c.f. Eqs~\eqref{eq:amres} and~\eqref{eq:ZAeff}) from which these quantities can
be determined at late times. As expected from our previous
work~\cite{Boyle:2016imm}, we find that for large quark masses, the residual
mass grows and eventually becomes unbounded. We conservatively discard any data
where this might be the case and only show data points for which the residual
mass reaches a plateau at late times. We observe stable plateaus for all data
points that are included in the analysis. Since the data is very precise and the
plateaus are unambiguous, we do not perform fits to the data, but simply take
the midpoint value (rightmost points in the plots in
Figure~\ref{fig:ZA_amres_M1M}). Numerical values for all data points are
presented in Tables~\ref{tab:dataC1M}-\ref{tab:dataF1S} in
Appendix~\ref{app:numericalresults}. Figure~\ref{fig:allmasses} shows the
spectrum as a function of the bare quark mass $m$.  Combining the determination
of $Z_A$ for each simulated mass point with the system of equations
Eqs.~\eqref{eq:Z_m}-\eqref{eq:Z_S} we obtain the corresponding values of
$Z^\msmom_m$ at each mass point for the simulated renormalization scales $\mu$.

\begin{table}
  \caption{Comparison of observables between the M0M ($M_\pi=139\,\mathrm{MeV}$)
    and the M1M ($M_\pi=286\,\mathrm{MeV}$) ensembles for two mass points
    bracketing the physical charm quark mass.}
  \label{tab:chiral}
  \begin{tabular}{c|lll}
    \hline\hline
    0.32 & M0M & M1M & M0M/M1M\\\hline
    $aM$ & 1.23636(19) & 1.23593(61) & 1.00035(53)\\
    $am_\mathrm{res}$ & 0.0006613(18) &  0.0006617(21) & 0.9993(41)\\
    $Z_A$ & 0.824110(43) & 0.824154(95) & 0.99995(12)\\\hline\hline
    0.34 & M0M & M1M & M0M/M1M\\\hline
    $aM$ & 1.28092(18) & 1.28049(61) & 1.00033(50)\\
    $am_\mathrm{res}$ & 0.0009049(26) & 0.0009004(28) & 1.0050(43)\\
    $Z_A$ & 0.833863(42) & 0.833897(100) & 0.99996(13)\\\hline\hline
  \end{tabular}
\end{table}
Before performing the required interpolations and continuum extrapolations, we
consider the size of potential effects afflicting simulations which do not take
place at the physical pion mass. Table ~\ref{tab:chiral} contrasts the values
for $aM$, $am_\mathrm{res}$ and $Z_A$ on the M0M and the M1M ensembles for two
choices of the heavy quark mass that bracket the physical charm quark
mass. These two ensembles only differ in their volume and pion mass. We observe
that the respective values on M0M and M1M are compatible with each other and
hence their ratios are compatible with unity. We further observe that the
relative (albeit not statistically resolved) effect on the hadron mass is at the
sub-permille level. We therefore conclude that any chiral effects in the data
can be safely neglected.

\subsection{Interpolations \label{subsec:interp}}
Having obtained $Z^\msmom_m(g,a\mu,am)$, $M(am)$ and $am$ at each simulated mass
point, we now perform the interpolations listed in steps~\ref{step1}
and~\ref{step2}.  Given the broad range covered by our data (cf. Figs
~\ref{fig:Zminterp} and \ref{fig:allmasses}), we perform these interpolations
locally as polynomial fits to the closest data points. In order to estimate any
systematic uncertainties stemming from these interpolations we perform them in
multiple ways:
\begin{itemize}
\item linear interpolation between the two closest bracketing datapoints.
\item quadratic interpolations between the two data points which bracket the
  target value and the nearest other data point to the left (right).
\item a cubic interpolation between the closest 4 points.
\end{itemize}
We take the quadratic interpolation with the third data point closest to the
target value as our central value and in addition to its statistical uncertainty
assign half the spread between these values as a systematic
uncertainty. Figure~\ref{fig:Zminterp} illustrates this for step~\ref{step1},
i.e. the interpolation of $Z_m$ at fixed mass ($am_q=0.15$) to the scale of
$\mu=2\,\mathrm{GeV}$). Since we want to contrast the approach to the continuum
limit between the massless (\smom) and the massive (\msmom) scheme, we also
compute $Z_m^\smom$.

For completeness, we list the numerical values for $Z^\msmom_m$ and $Z_m^\smom$
at $\mu=2\,\mathrm{GeV}$ in Tables~\ref{tab:dataC1M}-\ref{tab:dataF1S}. We
compute $Z_m^\smom$ only in the light and strange sector and use these values to
extrapolate $Z_m^\smom$ to the massless limit, prior to applying it.

\subsection{Continuum extrapolations \label{subsec:CL}}

\begin{figure}
  \includegraphics[width=\columnwidth]{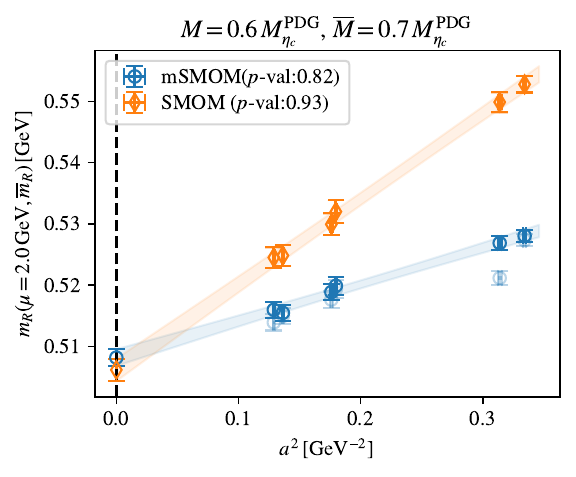}
  \includegraphics[width=\columnwidth]{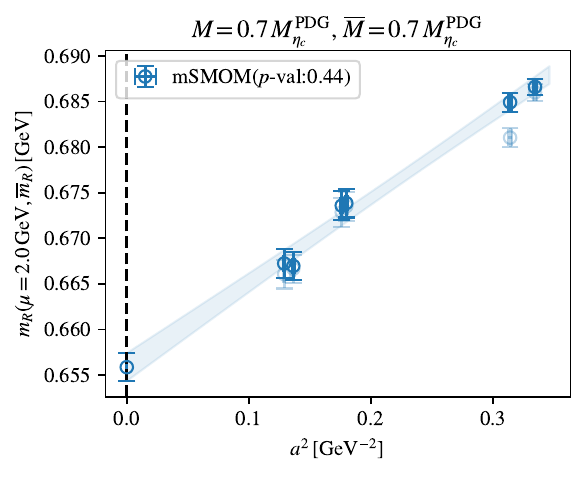}
  \caption{\emph{Top}: Example continuum limit extrapolation comparing the
    approach to the continuum for the SMOM scheme to the that of the mSMOM
    scheme. The quark masses $m_R$ (that is being extrapolated) and
    $\overline{m}_R$ (defining the \msmom scheme) are chosen to reproduce mesons
    of mass $0.6\times M_{\eta_c}^\mathrm{PDG}$ and $0.7\times
    M_{\eta_c}^\mathrm{PDG}$, respectively. \emph{Bottom}: Continuum limit
    determining the renormalization mass scale $\overline{m}_R$ at which the
    renormalization conditions are imposed.}
  \label{fig:CLexpl}
\end{figure}

Having determined $am$, $a\overline{m}$ and $Z_m(g,a\hat{\mu},a\overline{m})$ on
each ensemble, we can now perform the continuum limit of the renormalized quark
mass $am$ using the \msmom scheme at a renormalization scale $\hat{\mu}$ and
mass scale $\overline{m}$. The most general ansatz that we consider for our
continuum extrapolations is given by
\begin{equation}
  m^X(a\Lambda,a\hat{\mu}) = m^X(\hat{\mu}) + C_\chi am_\mathrm{res} + C_1 (a\Lambda)^2 \,,
  \label{eq:CLansatz}
\end{equation}
where the coefficient $C_\chi$ captures scaling violations stemming from the
residual chiral symmetry breaking in our data. We tried adding a term
proportional to $a^4$ but in practice we find that the term proportional to
$a^4$ is compatible with zero and not needed to describe the data and we hence
do not include it in the ansatz. Contrary to this, the coefficient $C_\chi$ is
typically resolved from zero and tends to be of $O(1)$. However, the size of
$am_\mathrm{res}$ is typically small
(c.f. Tables~\ref{tab:dataC1M}-\ref{tab:dataF1S}).

We present an example continuum limit fit in the top panel of
Figure~\ref{fig:CLexpl} for the choice $\overline{M} = 0.7 \times
M_{\eta_c}^\mathrm{PDG}$ and $M = 0.6 \times M_{\eta_c}^\mathrm{PDG}$. In
addition to the \msmom data points (blue circles) we also show the approach to
the continuum limit using the chirally extrapolated value of $Z_m$ in the \smom
scheme (orange diamonds). We clearly see that the data has smaller
discretisation effects in the \msmom scheme than the \smom scheme. The continuum
extrapolated values are not expected to agree with each other, since they are
not converted to the same scheme yet. However, when evaluating the conversion
factors for the scale and mass at hand (compare the right hand panel of
Fig~\ref{fig:mass-conv-factor} in Appendix~\ref{app:mass-conversion}), we find
the conversion factor to be very close to unity. In order to determine the exact
parameters of the scheme it remains to determine the value of $\overline{m}$,
i.e. to take the continuum limit where $M = \overline{M}$. This is shown for the
value $0.7 \times M_{\eta_c}^\mathrm{PDG}$ in the bottom panel of
Figure~\ref{fig:CLexpl}.

\begin{figure}
  \includegraphics[width=\columnwidth]{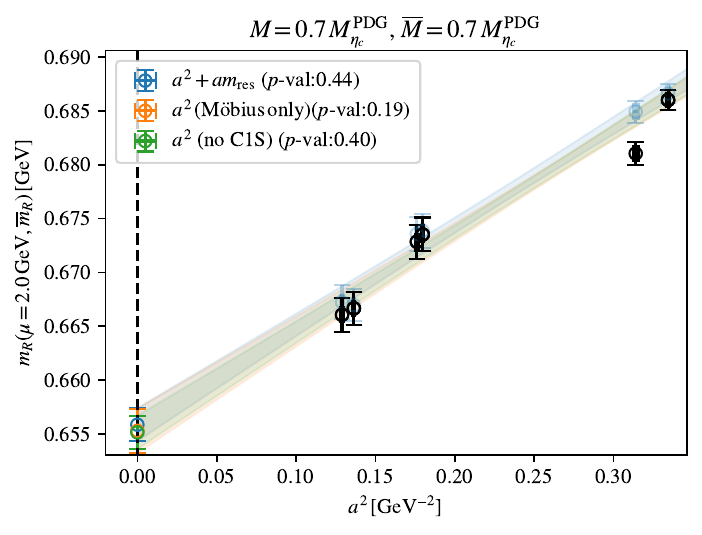}
  \caption{Variations of the continuum limit extrapolation presented in the top
    panel of Fig.~\ref{fig:CLexpl}}
  \label{fig:CLsys}
\end{figure}

In both plots, the original \msmom data points are shown as partially
transparent blue symbols, the opaque blue symbols present the value once the
residual mass contribution is corrected for. We notice that this only
significantly affects the C1S data point, which is expected since residual
chiral symmetry breaking effects are known to decrease as the lattice spacing is
reduced and when increasing the M\"obius scale which is one for the Shamir
kernel and two for the M\"obius kernel we are using.\footnote{The residual
  chiral symmetry breaking of our choice of M\"obius kernel is expected to be
  the same as that of the Shamir kernel with twice the extent of the fifth
  dimension $L_s$. Since $L_s(C1S) = 16$ and $L_s(C1M) = 24$ the C1M ensemble
  effectively has a three times larger extent of the fifth dimension.} In
addition, the residual mass is known to increase as the input quark mass $am_q$
increases as can be seen e.g. in the top panel of Fig~\ref{fig:ZA_amres_M1M}.

In order to assess the systematic uncertainties associated to the continuum
limit extrapolation we repeat the fit for several variations. In particular
Figure~\ref{fig:CLsys} shows this for the case of the combination of masses
$(M,\overline{M})$ presented in the top panel of Fig.~\ref{fig:CLexpl}. We
consider
\begin{itemize}
\item fitting all ensembles on which the hadron mass $M$ can be simulated
  including the terms proportional to $am_\mathrm{res}$ and $(a\Lambda)^2$
\item fitting all except the C1S ensemble (which has by far the largest
  $am_\mathrm{res}$ value) only including the term proportional to
  $(a\Lambda)^2$
\item fitting only the M\"obius ensembles (which have smaller $am_\mathrm{res}$
  values) only including the term proportional to $(a\Lambda)^2$.
\end{itemize}
We quote the first of these fits as our central value and additionally assign
half the spread of the variations as a systematic from the choice of continuum
limit.

\subsection{Varying the renormalization scales}

\begin{figure}
  \includegraphics[width=\columnwidth]{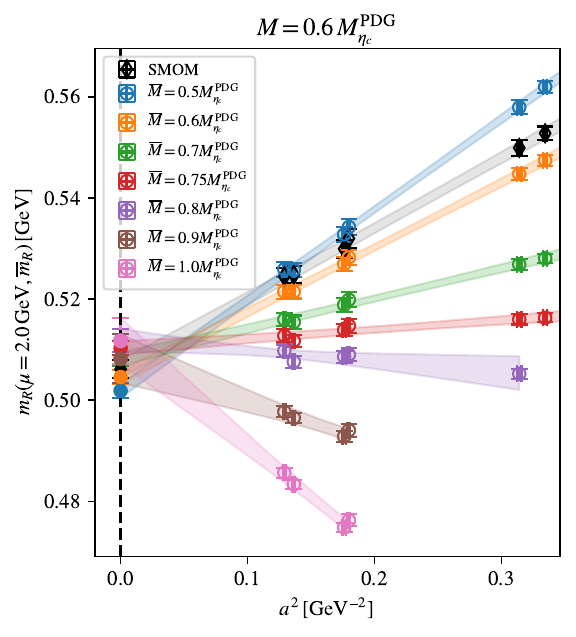}
  \caption{Variations of the renormalization mass scale $\overline{M}$ at fixed
    value of $M$. The data is fitted to Eq.~\eqref{eq:CLansatz} and the plot
    displays it after correcting to vanishing residual mass. The black data
    points show the approach to the continuum for the case of the massless \smom
    scheme. We stress that different values of $\overline{m}$ define different
    schemes, hence these numbers are \emph{not} expected to agree in the
    continuum limit.}
  \label{fig:Mbar-variation}
\end{figure}

We stress that the renormalization mass scale $\overline{m}_R$ set by
$\overline{M}$ is a scale that can be varied freely within the range where we
have data. In Figure~\ref{fig:Mbar-variation} we presents fits to the ansatz
Eq.~\eqref{eq:CLansatz} $M = 0.6 \times M_{\eta_c}^\mathrm{PDG}$ but for a
variety of choices of $\overline{M}$.  We emphasise again, that the extrapolated
values do not have to agree as they are still in different schemes. It is
however clearly visible that the approach to the continuum is well described by
a fit linear in $a^2$ but that the slope varies strongly with the choice of
$\overline{M}$. For the largest values of $\overline{M}$ we lose coverage on the
coarsest ensembles and hence remove them from the fit. For the remaining
analysis we restrict ourselves to values of $\overline{M}$ that allow direct
simulations on all considered ensembles.

We also vary the renormalization scale $\mu$ between $2.0\,\mathrm{GeV}$,
$2.5\,\mathrm{GeV}$ and $3.0\,\mathrm{GeV}$. We observe that for increasing
values of $\mu$ the values of $\overline{M}$ which are required to significantly
flatten the continuum limit are beyond the range where we can determine
$\overline{m}$ from all three lattice spacings. We therefore base our final
results on continuum limit extrapolations at $\mu=2\,\mathrm{GeV}$ and only show
the corresponding \msmom results obtained from larger scales for comparison
(c.f. Table~\ref{tab:results}).

\subsection{The charm quark mass}

\begin{figure}
  \includegraphics[width=\columnwidth]{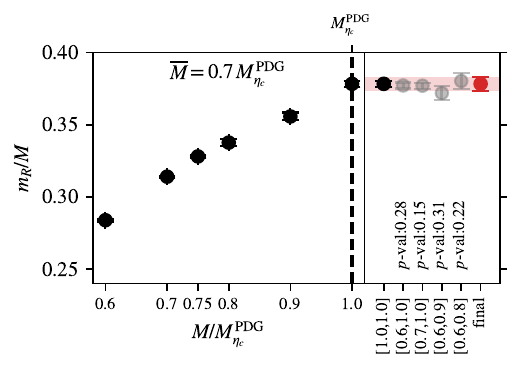}
  \caption{Extrapolation of different intervals of the various reference values
    $M$ to the physical value $M_{\eta_c}^\mathrm{PDG}$ at fixed
    $\overline{M}$. The right hand panel shows results for these
    interval-choices. The red datapoint to the very right and the corresponding
    band represents our final value of $m_{c,R}$ in the \msmom scheme at
    $\overline{M}=0.7\,M_{\eta_c}^\mathrm{PDG}$.}
  \label{fig:charm-extrap}
\end{figure}

We now vary the choice of $M$ using the various $M_i$ and repeat the continuum
limit fit for each case, keeping $\overline{M}$ fixed in order to remain in the
same scheme. For each choice $M_i$ we assemble the error budget of this fit to
obtain values $m^{\msmom}_i(2\,\mathrm{GeV},\overline{m})$. We now combine these
results to perform an inter- or extrapolation to the physical charm-quark
mass. This is not strictly speaking necessary, since we already have a direct
result for this quark mass from the continuum limit at $M_i =
M_{\eta_c}^\mathrm{PDG}$, however since this continuum limit is only based on
the medium and fine ensembles we prefer to supplement it by a parameterisation
using different values (in the continuum) of $m_i^R/M_i$ as a function of
$M_i$. This is shown in Figure~\ref{fig:charm-extrap}. Of our choices for $M_i$
we consider the ranges $M_i/M_{\eta_c}^\mathrm{PDG}$ $[0.6,1.0]$, $[0.7,1.0]$,
$[0.6,0.9]$ and $[0.6,0.8]$. In each case we parameterise the dependence of the
quark mass as a polynomial in $M_i$ via
\begin{equation}
  \frac{m_R}{M} = \alpha M^{-1} + \beta  + \gamma M\,.
  \label{eq:charmmassextrap}
\end{equation}

\begin{table*}
  \caption{Summary of our final results for the charm quark mass. We quote the
    renormalization mass scale $\overline{m}_R^\msmom$ defining the given \msmom
    scheme, as well as the charm quark mass in that scheme, converted to $\ms$
    at the same scale and to $\ms$ at $\overline{m}_c$. All quark masses are
    given in units of $\mathrm{GeV}$.}
  \label{tab:results}
  \begin{tabular}{cccclll}
    \hline\hline
    $\mu/\mathrm{GeV}$ & $\overline{M}/M_{\eta_c}^\mathrm{PDG}$ & $\overline{m}_R(\mu,\overline{m}_R)$  & $m^{\mathrm{RI}}_{c,R}(\mu,\overline{m}_R)$ &   $\overline{m}_c^{\overline{\mathrm{MS}}}(\mu)$ & $\overline{m}_c^{\overline{\mathrm{MS}}}(3\,\mathrm{GeV}\leftarrow \mu)$ & $\overline{m}_c^{\overline{\mathrm{MS}}}(\overline{m}_c)$ \\\hline
    2.0 & 0.60 & 0.5046(15) & 1.127(7)(12) & 1.112(7)(12)(4) & 1.005(6)(11)(4) & 1.289(6)(10)(3) \\
    2.0 & 0.70 & 0.6559(16) & 1.129(7)(12) & 1.115(7)(12)(4) & 1.008(6)(11)(4) & 1.292(5)(10)(4)\\
    2.0 & 0.75 & 0.7371(16) & 1.130(6)(13) & 1.118(6)(13)(4) & 1.010(6)(11)(4) & 1.294(5)(10)(4) \\ 
    2.0 & SMOM & ---        & 1.136(9)(12) & 1.114(9)(12) & 1.007(8)(10) & 1.291(8)(10)\\
    \hline
    2.5 & 0.60 & 0.4698(14) & 1.052(7)(14) & 1.038(7)(14)(3) & 0.995(7)(14)(3) & 1.280(6)(13)(3) \\
    2.5 & 0.70 & 0.6124(16) & 1.057(6)(15) & 1.043(6)(15)(3) & 1.000(6)(14)(3) & 1.284(6)(13)(3) \\
    2.5 & 0.75 & 0.6894(16) & 1.059(6)(15) & 1.046(6)(15)(3) & 1.003(6)(15)(3) & 1.287(5)(13)(3) \\
    2.5 & SMOM & ---        & 1.066(11)(12)& 1.048(10)(12) & 1.004(10)(12) & 1.288(9)(11) \\
    \hline
    3.0 & 0.60 & 0.4450(14) & 0.998(7)(15) & 0.986(7)(15)(3) & 0.986(7)(15)(3) & 1.271(6)(14)(2) \\
    3.0 & 0.70 & 0.5811(15) & 1.004(6)(15) & 0.992(6)(15)(3) & 0.992(6)(15)(3) & 1.277(5)(14)(2) \\
    3.0 & 0.75 & 0.6549(16) & 1.008(6)(16) & 0.995(6)(16)(3) & 0.995(6)(16)(3) & 1.280(5)(15)(2) \\
    3.0 & SMOM & ---        & 1.018(8)(12) & 1.002(8)(12) & 1.002(8)(12) & 1.287(8)(11) \\
    \hline\hline
  \end{tabular}
\end{table*}
The result to these variations is shown in the right-hand panel of
Figure~\ref{fig:charm-extrap}. We take the direct determination at the charm
quark mass (i.e. $M_i = M_{\eta_c}^\mathrm{PDG}$) as central value and in
addition to its uncertainty we conservatively associate a systematic uncertainty
of half the spread of the variation. These fit results that determine this
uncertainty are shown in the right hand panel of
Figure~\ref{fig:charm-extrap}. We quote the two uncertainties separately, since
the latter only arises since we require our final number to be based on
continuum limits from more than two lattice spacings. With additional finer
ensembles, this last uncertainty would be completely removed, since a continuum
limit with three lattice spacings could be obtained directly at the charm quark
mass. We then repeat the analysis for different choices of $\overline{M}$ (and
hence $\overline{m}_R$) as well as the massless \smom scheme. 

Finally, it remains to convert these results into a common scheme where they can
be directly compared to each other. Using the conversion factor
$R_m^{\overline{\mathrm{MS}} \leftarrow \msmom}$ given in
Eq.~\eqref{eq:mSMOMtoMSbar}, we can convert the results from
$m_{c,R}^\msmom(\mu,\overline{m}_R)$ to $m_{c,R}^{\ms}(\mu)$. Unfortunately, for
the \msmom scheme, this is currently only known at one loop. In order to
quantify the truncation effects in the temporary absence of perturbative
two-loop calculations, we investigate the difference between one- and two-loop
corrections for the massless scheme~\cite{Gorbahn:2010bf, Almeida:2010ns} and
assign the relative difference between them as a systematic truncation
uncertainty. In practice we find that for $\mu=2.0\,\mathrm{GeV}$
($2.5\,\mathrm{GeV}$, $3.0\,\mathrm{GeV}$) the difference between 1- and 2-loop
conversion to \ms is a 0.38\% (0.31\%, 0.27\%) effect. Within the \ms scheme we
then run the results up to $3\,\mathrm{GeV}$ as well as down to the charm quark
scale to quote $\overline{m}_c(\overline{m}_c)$. To compute the strong coupling
and running of the \msbar quark mass we make use of
RunDec~\cite{Chetyrkin:2000yt, Schmidt:2012az, Herren:2017osy}, which in turn
relies on 5-loop results for the beta function and for the mass anomalous
dimension~\cite{Baikov:2014qja, Luthe:2016xec, Baikov:2017ujl, Baikov:2016tgj,
  Herzog:2017ohr, Luthe:2017ttc}.

\begin{figure}
  \includegraphics[width=\columnwidth]{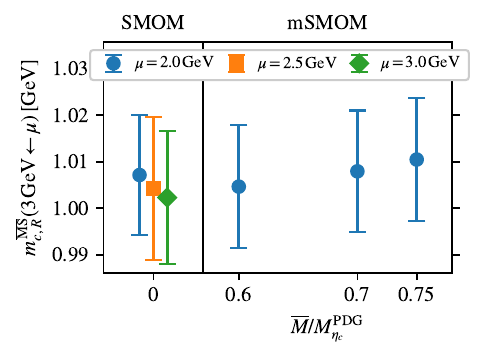}
  \caption{Results for the continuum-extrapolated renormalized charm quark mass
    converted to $\overline{\mathrm{MS}}$ from RI/SMOM and RI/mSMOM with
    variation in $\overline{M}$ at renormalization scale of 3 GeV using results
    from $\mu=2.0, 2.5, 3.0\,\mathrm{GeV}$. Numerical values are presented in
    Table~\ref{tab:results}.}
  \label{fig:varymbar}
\end{figure}

We list these results for some choices of $\overline{M}$ and $\mu$ in
Table~\ref{tab:results} where the first uncertainty is the result from the pure
\msmom calculation at at a chosen quark mass, the second uncertainty lists
encapsulates the charm-mass interpolation, and the last uncertainty estimates
the truncation effect due to performing the matching at 1-loop. In principle one
could also quote a fourth uncertainty encapsulating the uncertainties of the
inputs to the running and the truncation of the running factor, however these
are found to be negligible.

Our final results for the charm quark mass converted to \ms and then (where
necessary) ran to $3\,\mathrm{GeV}$ within the \ms scheme are shown in
Figure~\ref{fig:varymbar}. As mentioned above, in the massive scheme the
continuum limit is well controlled for determinations at $\mu=2$ but values of
$\overline{M}$ which significantly decrease the slope of the continuum limit are
not reachable on our current data set for larger values of $\mu$ and we
therefore exclude them. We find good agreement between the \smom and the \msmom
schemes as well as amongst different values for $\overline{m}_R$ within the
\msmom scheme. As our final number we quote our results obtained from \msmom at
$\hat{\mu}=2\,\mathrm{GeV}$ from the choice $\overline{M} =
0.7\,M_{\eta_c}^\mathrm{PDG}$ which corresponds to $\overline{m}_R^\msmom =
0.6559(16)\,\mathrm{GeV}$. We find
\begin{align}
  m^{\msmom}_{c,R}(2\,\mathrm{GeV},\overline{m}_R) &= 1.129(7)(12)\,\mathrm{GeV}\,,\\
  m^{\overline{\mathrm{MS}}}_{c,R}(2\,\mathrm{GeV}) &= 1.115(7)(12)(4)\,\mathrm{GeV}\,,\\
  m^{\overline{\mathrm{MS}}}_{c,R}(3\,\mathrm{GeV}) &= 1.008(6)(11)(4)\,\mathrm{GeV}\,,\\
  m^{\overline{\mathrm{MS}}}_{c,R}(m^{\overline{\mathrm{MS}}}_{c,R}) &= 1.292(5)(10)(4)\,\mathrm{GeV}\,.
\end{align}
The first uncertainty comes from the determination directly at the charm quark
mass, the second from the inter-/extrapolation taking smaller than physical
reference values, the third from the perturbative truncation uncertainty when
converting to \ms. We have not applied any additional uncertainties associated
with the running within \ms.

\subsection{Comparison to the literature}
\begin{figure}
  \includegraphics[width=\columnwidth]{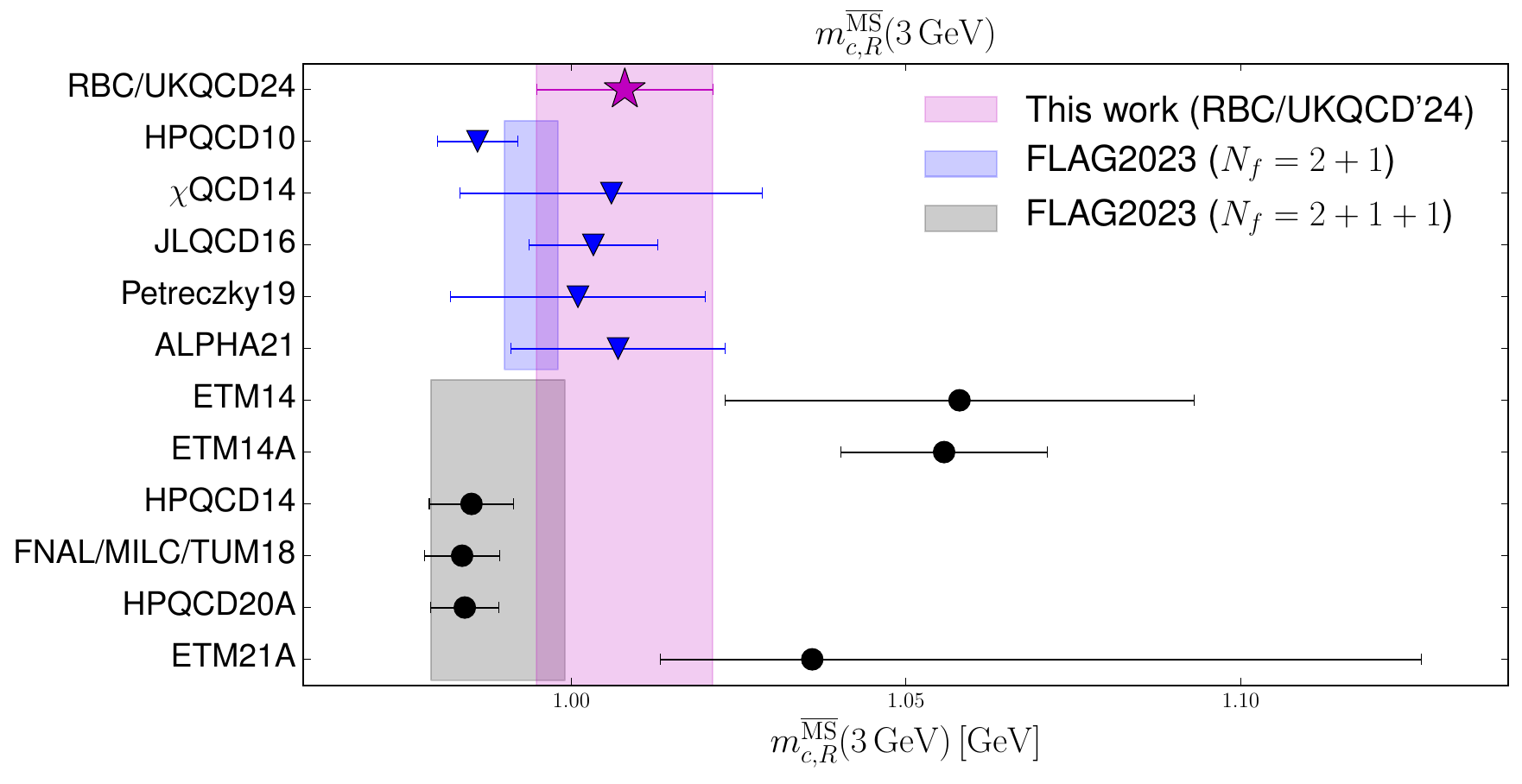}
  \caption{Comparison of our result with previous determinations of the charm
    quark mass in the literature that enter the
    FLAG~\cite{FlavourLatticeAveragingGroupFLAG:2021npn} average. The magenta
    star and corresponding band are our result, the blue triangles are results
    based on $N_f=2+1$~\cite{Heitger:2021apz, Petreczky:2019ozv,
      Nakayama:2016atf, Yang:2014sea, McNeile:2010ji} and the black circles
    based on $N_f=2+1+1$~\cite{ExtendedTwistedMass:2021gbo, Hatton:2020qhk,
      FermilabLattice:2018est, Alexandrou:2014sha, Chakraborty:2014aca,
      EuropeanTwistedMass:2014osg}. The blue and black bands show the
    corresponding FLAG averages for $N_f=2+1$ and $N_f=2+1+1$, respectively.}
  \label{fig:comp}
\end{figure}

The charm quark mass has been previously computed by various collaborations in
various schemes. In Figure~\ref{fig:comp} we compare our result to the results
in the literature which enter the FLAG average in the \ms scheme at
$3\,\mathrm{GeV}$\footnote{In cases where the results were quoted at
  $2\,\mathrm{GeV}$, we follow FLAG's convention and apply a running factor of
  0.900 to obtain the result at $3\,\mathrm{GeV}$.} We find good agreement with
other $N_f=2+1$ calculations and obtain similar uncertainties. The leading
uncertainty in our calculation arises from the fact that not all the ensembles we
currently use allow for direct simulation at the physical charm quark
mass. Using an additional finer lattice spacing will allow to eliminate this
uncertainty in the future.

\section{Summary and Outlook \label{sec:conc}}
We have presented the first numerical implementation of the massive
non-perturbative renormalization scheme which was first suggested in
Ref.~\cite{Boyle:2016wis}. We find that varying the mass scale at which the
renormalization conditions are imposed can be used to significantly modify the
approach to the continuum limit, and in particular to flatten it. We observe
good agreement between different renormalization mass scales (and hence
continuum limit approaches), further substantiating that the continuum limit is
controlled.

This scheme can be applied to any observable and hence be used to obtain more
reliable continuum limits. Since different choices of the renormalization mass
scale must agree in the continuum limit, it also provides non-trivial tests
allowing to scrutinise a given continuum limit by performing it for different
choices of the renormalization scheme mass scale. This is not restricted to
heavy-quark observables but is useful for any observable with large
discretisation effects compared to the desired statistical precision.

The joint continuum limit fits to the chosen M\"obius und Shamir domain wall
kernels with very similar lattice spacings agree well with only fitting the
M\"obius ensembles and (at the present level of precision) are well described by
an ansatz that is linear in $a^2$. We obtain the charm quark mass in the \ms
scheme at $3\,\mathrm{GeV}$ with a precision of 1.3\% in good agreement with the
literature. This uncertainty can be significantly reduced by using additional
finer ensembles. By direct computation, we find the sea-pion effect on the
charmed meson mass $M_{\eta_c}$ between a pion mass of $286\,\mathrm{MeV}$ and
the physical pion mass to be below the permille level.

For the future, we envisage applications of the \msmom scheme to other
observables and an extension to four quark operators.
\section*{Acknowledgements}
We thank our colleagues in the RBC and UKQCD collaborations for many fruitful
discussions.  We are particularly grateful to Peter Boyle for valuable
discussions and maintaining the Grid software package. We thank Chris Sachrajda
and Matteo Di Carlo for helpful theoretical discussions in the early stages of
this work. We thank Antonin Portelli for maintaining the Hadrons package.

L.D.D.\ is funded by the UK Science and Technology Facility Council (STFC) grant
ST/P000630/1 and by the ExaTEPP project EP/X01696X/1.  F.E.\ has received
funding from the European Union’s Horizon Europe research and innovation
programme under the Marie Sk\l{}odowska-Curie grant agreement
No.\ 101106913. R.M.\ is funded by a University of Southampton Presidential
Scholarship. This work used the DiRAC Extreme Scaling service at the University
of Edinburgh, operated by the Edinburgh Parallel Computing Centre on behalf of
the STFC DiRAC HPC Facility (www.dirac.ac.uk). This equipment was funded by BIS
National E-infrastructure capital grant ST/K000411/1, STFC capital grant
ST/H008845/1, and STFC DiRAC Operations grants ST/K005804/1 and
ST/K005790/1. DiRAC is part of the National e-Infrastructure.

\appendix

\section{Numerical results}\label{app:numericalresults}
In tables~\ref{tab:dataC1M}-~\ref{tab:dataF1S} we summarise the numerical data
for the residual mass, $Z_A$, the hadron mass $aM$ as well as the
renormalization constant $Z_m$ interpolated to a scale of $2\,\mathrm{GeV}$.

\begin{table}
  \caption{Summary of numerical results on the C1M ensemble used for the
    analysis. The values of $Z_m$ are given after interpolation to $\mu =
    2\,\mathrm{GeV}$.}  \resizebox{\columnwidth}{!}{ \begin{tabular}{l|lllllll}

C1M & $10^3 am_\mathrm{res}$ & $aM$ & $Z_A$ & $Z^\mathrm{mSMOM}_m$ & $Z_m^{\mathrm{SMOM}}$\\ \hline
0.0050&  0.601(12) &  0.1642(34) &  0.71302(34) &  1.5754(65) &  1.5428(15) \\
0.0100&  0.574(11) &  0.2203(22) &  0.71337(19) &  1.6088(36) &  1.5442(12) \\
0.0181&  0.5330(95) &  0.2886(14) &  0.71443(12) &  1.6211(22) &  1.5425(11) \\
0.0362&  0.4642(79) &  0.40331(87) &  0.717257(77) &  1.6232(15) &  1.5362(10) \\
0.0500&  0.450(17) &  0.4769(22) &  0.71979(19) &  1.6185(14) & - \\
0.1000&  0.361(12) &  0.6877(14) &  0.72921(11) &  1.5996(12) & - \\
0.1500&  0.3210(100) &  0.8637(11) &  0.74087(11) &  1.5725(10) & - \\
0.2000&  0.3172(94) &  1.01971(85) &  0.75528(13) &  1.53813(93) & - \\
0.3000&  0.599(16) &  1.28930(50) &  0.79647(16) &  1.44377(74) & - \\

\end{tabular}
 }
   \label{tab:dataC1M}
\end{table}

\begin{table}
  \caption{Summary of numerical results on the C1S ensemble used for the
    analysis. The values of $Z_m$ are given after interpolation to $\mu =
    2\,\mathrm{GeV}$.}  \resizebox{\columnwidth}{!}{ \begin{tabular}{l|lllllll}

C1S & $10^3 am_\mathrm{res}$ & $aM$ & $Z_A$ & $Z^\mathrm{mSMOM}_m$ & $Z_m^{\mathrm{SMOM}}$\\ \hline
0.0050&  3.162(19) &  0.1885(29) &  0.71796(44) &  1.3145(75) &  1.5410(19) \\
0.0100&  3.085(18) &  0.2367(21) &  0.71822(28) &  1.4381(35) &  1.5411(14) \\
0.0200&  2.938(16) &  0.3129(15) &  0.71942(17) &  1.5245(18) &  1.5392(12) \\
0.0400&  2.720(12) &  0.4304(10) &  0.72235(12) &  1.5706(12) &  1.5325(12) \\
0.0500&  2.644(22) &  0.4829(18) &  0.72413(27) &  1.5772(14) & - \\
0.1000&  2.427(15) &  0.6904(14) &  0.73344(23) &  1.5763(12) & - \\
0.1500&  2.420(11) &  0.8636(11) &  0.74507(19) &  1.5542(11) & - \\
0.2000&  2.6192(92) &  1.01733(94) &  0.75960(15) &  1.5210(10) & - \\
0.3000&  4.530(18) &  1.28409(79) &  0.80194(14) &  1.42069(78) & - \\
0.3300&  6.455(18) &  1.35527(38) &  0.821474(90) &  1.37464(67) & - \\

\end{tabular}
 }
  \label{tab:dataC1S}
\end{table}

\begin{table}
  \caption{Summary of numerical results on the M1M ensemble used for the
    analysis. The values of $Z_m$ are given after interpolation to $\mu =
    2\,\mathrm{GeV}$.}  \resizebox{\columnwidth}{!}{ \begin{tabular}{l|lllllll}

M1M & $10^3 am_\mathrm{res}$ & $aM$ & $Z_A$ & $Z^\mathrm{mSMOM}_m$ & $Z_m^{\mathrm{SMOM}}$\\ \hline
0.0040&  0.3116(61) &  0.1196(26) &  0.74376(24) &  1.5167(51) &  1.5743(21) \\
0.0080&  0.3018(56) &  0.1651(16) &  0.74421(13) &  1.5635(26) &  1.5722(23) \\
0.0133&  0.2907(51) &  0.2113(12) &  0.744798(86) &  1.5820(20) &  1.5709(20) \\
0.0266&  0.2709(39) &  0.29939(79) &  0.746330(56) &  1.5955(18) &  1.5667(18) \\
0.0500&  0.2527(54) &  0.4178(17) &  0.749495(88) &  1.5970(16) & - \\
0.1000&  0.2414(40) &  0.6163(11) &  0.757548(60) &  1.5851(15) & - \\
0.1500&  0.2523(35) &  0.78311(79) &  0.767843(47) &  1.5612(14) & - \\
0.2250&  0.3173(27) &  1.00082(63) &  0.788084(41) &  1.5093(11) & - \\
0.3000&  0.5277(20) &  1.19017(57) &  0.815321(43) &  1.44227(90) & - \\
0.3200&  0.6634(21) &  1.23610(64) &  0.824062(46) &  1.42193(84) & - \\
0.3400&  0.8998(26) &  1.28063(63) &  0.833810(56) &  1.39986(79) & - \\

\end{tabular}
 }
  \label{tab:dataM1M}
\end{table}

\begin{table}
  \caption{Summary of numerical results on the M1S ensemble used for the
    analysis. The values of $Z_m$ are given after interpolation to $\mu =
    2\,\mathrm{GeV}$.}  \resizebox{\columnwidth}{!}{ \begin{tabular}{l|lllllll}

M1S & $10^3 am_\mathrm{res}$ & $aM$ & $Z_A$ & $Z^\mathrm{mSMOM}_m$ & $Z_m^{\mathrm{SMOM}}$\\ \hline
0.0040&  0.6727(72) &  0.1290(24) &  0.74486(27) &  1.5055(58) &  1.5720(26) \\
0.0080&  0.6561(64) &  0.1714(16) &  0.74542(15) &  1.5557(28) &  1.5708(21) \\
0.0150&  0.6319(55) &  0.2283(11) &  0.746212(91) &  1.5802(19) &  1.5690(19) \\
0.0300&  0.5977(42) &  0.32118(69) &  0.747987(58) &  1.5930(18) &  1.5639(18) \\
0.0500&  0.5767(52) &  0.42023(76) &  0.75062(11) &  1.5951(18) & - \\
0.1000&  0.5479(35) &  0.61780(44) &  0.758721(100) &  1.5838(17) & - \\
0.1500&  0.5602(29) &  0.78382(43) &  0.769158(89) &  1.5596(15) & - \\
0.2250&  0.6677(29) &  1.00056(52) &  0.789767(76) &  1.5066(13) & - \\
0.3000&  1.0409(41) &  1.18880(56) &  0.817587(74) &  1.43775(99) & - \\
0.3200&  1.2562(65) &  1.23385(41) &  0.826456(79) &  1.41695(97) & - \\
0.3400&  1.6053(82) &  1.27801(40) &  0.836317(81) &  1.39437(87) & - \\
0.3600&  2.189(11) &  1.32043(39) &  0.847392(85) &  1.36951(81) & - \\
0.3750&  2.936(12) &  1.35187(55) &  0.857047(90) &  1.34809(77) & - \\

\end{tabular}
 }
  \label{tab:dataM1S}
\end{table}

\begin{table}
  \caption{Summary of numerical results on the F1M ensemble used for the
    analysis. The values of $Z_m$ are given after interpolation to $\mu =
    2\,\mathrm{GeV}$.}  \resizebox{\columnwidth}{!}{ \begin{tabular}{l|lllllll}

F1M & $10^3 am_\mathrm{res}$ & $aM$ & $Z_A$ & $Z^\mathrm{mSMOM}_m$ & $Z_m^{\mathrm{SMOM}}$\\ \hline
0.0021&  0.2399(56) &  0.0865(21) &  0.75927(21) &  1.4816(57) &  1.5797(23) \\
0.0043&  0.2390(52) &  0.1172(16) &  0.75952(11) &  1.5229(31) &  1.5802(21) \\
0.0107&  0.2343(43) &  0.1795(10) &  0.760226(53) &  1.5766(21) &  1.5792(19) \\
0.0214&  0.2286(36) &  0.25287(54) &  0.761281(42) &  1.5924(20) &  1.5759(18) \\
0.0330&  0.2244(31) &  0.31620(38) &  0.762536(41) &  1.5942(19) & - \\
0.0660&  0.2201(21) &  0.46183(32) &  0.766829(40) &  1.5935(19) & - \\
0.0990&  0.2248(15) &  0.58391(32) &  0.772132(39) &  1.5838(18) & - \\
0.1320&  0.2378(12) &  0.69368(30) &  0.778456(38) &  1.5683(16) & - \\
0.1980&  0.29064(77) &  0.88979(25) &  0.794371(35) &  1.5243(14) & - \\
0.2640&  0.39970(57) &  1.06271(21) &  0.815121(32) &  1.4682(11) & - \\
0.3300&  0.66808(62) &  1.21606(19) &  0.841614(31) &  1.40308(87) & - \\
0.3600&  1.0280(12) &  1.27967(19) &  0.856367(32) &  1.36962(80) & - \\

\end{tabular}

  }
  \label{tab:dataF1M}
\end{table}

\begin{table}
  \caption{Summary of numerical results on the F1S ensemble used for the
    analysis. The values of $Z_m$ are given after interpolation to $\mu =
    2\,\mathrm{GeV}$.} \resizebox{\columnwidth}{!}{ \begin{tabular}{l|lllllll}

F1S & $10^3 am_\mathrm{res}$ & $aM$ & $Z_A$ & $Z^\mathrm{mSMOM}_m$ & $Z_m^{\mathrm{SMOM}}$\\ \hline
0.0021&  0.9769(95) &  0.0994(18) &  0.76231(18) &  1.4979(67) &  1.5802(18) \\
0.0043&  0.9722(88) &  0.1263(13) &  0.76263(11) &  1.5265(38) &  1.5809(18) \\
0.0107&  0.9565(62) &  0.18387(86) &  0.763139(55) &  1.5759(23) &  1.5802(17) \\
0.0214&  0.9393(43) &  0.25453(57) &  0.764180(44) &  1.5918(19) &  1.5768(17) \\
0.0330&  0.9291(36) &  0.31626(41) &  0.765486(42) &  1.5935(20) & - \\
0.0660&  0.9188(24) &  0.45975(34) &  0.769873(38) &  1.5923(19) & - \\
0.0990&  0.9251(18) &  0.58074(34) &  0.775231(37) &  1.5822(17) & - \\
0.1320&  0.9463(14) &  0.68965(34) &  0.781619(36) &  1.5662(16) & - \\
0.1980&  1.0427(11) &  0.88429(32) &  0.797730(37) &  1.5208(13) & - \\
0.2640&  1.2577(11) &  1.05584(28) &  0.818823(38) &  1.4630(10) & - \\
0.3300&  1.7485(18) &  1.20763(25) &  0.845736(38) &  1.39618(82) & - \\
0.3960&  3.1873(46) &  1.34062(22) &  0.880386(40) &  1.31942(65) & - \\

\end{tabular}
 }
  \label{tab:dataF1S}
\end{table}

\section{Mass conversion factor}
\label{app:mass-conversion}

\begin{figure*}
 \begin{center}
  \includegraphics[width=\textwidth]{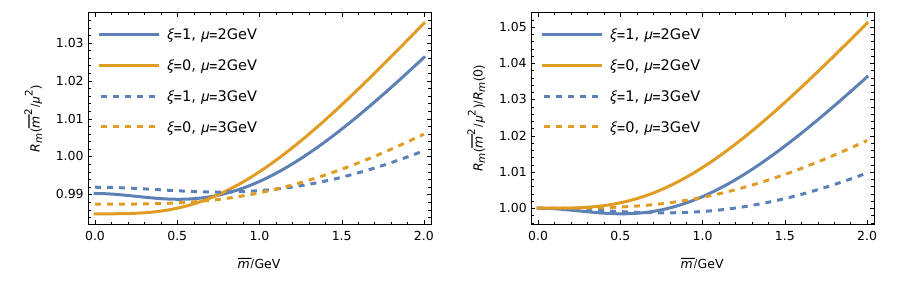}
 \end{center}
 \caption{Left: conversion factor $R_m^{\msbar\leftarrow
      \text{mSMOM}}(m^2/\mu^2)$ for $0<m<2\,\text{GeV}$. Right, ratio
    $R_m^{\msbar\leftarrow \text{mSMOM}}(m^2/\mu^2)/R_m^{\msbar\leftarrow\text{mSMOM}}(0) =
    R_m^{\text{SMOM}\leftarrow\text{mSMOM}}(m^2/\mu^2)$. Both plots show
    curves with $(\alpha,\mu)=(0.293347, 2\,\text{GeV})$ (solid) and
    $(\alpha,\mu)=(0.24358, 3\,\text{GeV})$ (dashed), and for
    gauge-parameter $\xi$ set to $1$ (Feynman gauge) or $0$ (Landau
    gauge). The dimensional regularisation scale $\tilde\mu$ is set to
    $\mu$.}
  \label{fig:mass-conv-factor}
\end{figure*}

Here we determine the mSMOM to $\msbar$ conversion factor for the mass
at one-loop in continuum perturbation theory, with an arbitrary choice
of the gauge parameter $\xi$. We work in Minkowski space, with fermion
propagator
\begin{equation}
S(p) = \frac i{\pslash - m - \Sigma(p) +i\eps}\,.
\end{equation}
We use dimensional regularisation in $d=4-2\eps$ dimensions, denoting
by $\tilde\mu$ the dimensionful scale introduced (to distinguish it
from the scale $\mu$ defining the mSMOM symmetric-momentum point). We
compute the fermion self-energy to find the wave-function
renormalization and then determine $\Zmmsmom$ by using the mSMOM
renormalization from equation~(\ref{eq:Z_m}), which in Minkowski space
reads
\begin{equation}
  \label{eq:Z_m-minkowski}
  \begin{aligned}
  1 &= \frac1{12m_R}\Big\{\Tr[-iS_R(p)^{-1}]_{p^2=-\mu^2}\\
  &\phantom{= \frac1{12m_R}}
    {}\;\;\; -\frac12 \Tr[q{\cdot}\Lambda_{\text{A},R}\gamma_5]_\text{sym}
  \Big\}.
  \end{aligned}
\end{equation}
The conversion factor is given by
\begin{equation}
  R_m^{\msbar\leftarrow \text{mSMOM}} = Z_m^\msbar/\Zmmsmom.
\end{equation}
The one-loop self-energy integral is
\begin{widetext}
\begin{equation}
  -i \Sigma^{(1)}(p) = -g^2\tilde\mu^{2\eps}C_F
  \int \frac{d^dk}{(2\pi)^d}\left[
  \frac{\gamma_\mu(\pslash -\kslash +m)\gamma^\mu}
       {k^2\big( (p-k)^2-m^2\big)}
  -(1-\xi)
  \frac{\kslash(\pslash -\kslash +m)\kslash}
       {(k^2)^2\big( (p-k)^2-m^2\big)}
  \right],
\end{equation}
where $C_F=4/3$ is the $\text{SU}(3)$ quadratic Casimir operator in
the fundamental representation. This can be evaluated by standard
techniques, using Mathematica~\cite{Mathematica} to perform the
Feynman-parameter integrals. The result is
\begin{equation}
  \label{eq:sigmap-xi}
  \begin{aligned}
  \Sigma^{(1)}(p) &= \afpC\!\! \left[
    \pslash\, \xi\left(-\frac1\epsbar - 1 + u
    + u^2\ln\Big(\!\frac u{1+u}\!\Big) +
    \ln(1+u) + \ln(\mu^2/\tilde\mu^2)\right)
    \right.\\
    & \phantom{= \afpC} \left.
     {} + m \left(4+2\xi +
     (3+\xi)\Big(\frac1\epsbar
    + u \ln\Big(\!\frac u{1+u}\!\Big)
    - \ln(1+u)- \ln(\mu^2/\tilde\mu^2\Big) \right)
    \right]\\
   &= \afpC \, [ \pslash A_\xi + m B_\xi ],
  \end{aligned}
\end{equation}
where we have set $u=m^2/\mu^2$ and have defined
\begin{equation}
  \frac1\epsbar \equiv \frac1\eps - \gamma_\text{E} + \ln(4\pi).
\end{equation}
When $\xi=1$ the result for $\Sigma^{(1)}(p)$ above agrees with
equation~(37) in Ref.~\cite{Boyle:2016wis}. From this, the mSMOM
wave-function renormalization constant, up to one loop, is
\begin{equation}
  \label{eq:Zqmsmom-xi}
  \Zqmsmom = 1 - \afpC\,A_\xi.
\end{equation}
Now write the one-loop contribution to an amputated bilinear vertex as
$\Lambda^{(1)}_\Gamma = \Lambda^{(1)}_{\Gamma,\xi=1} +
\Lambda^{(1)}_{\Gamma,\xi\neq1}$, with:
\begin{align}
  \Lambda^{(1)}_{\Gamma,\xi=1} &= 
  -ig^2\tilde\mu^{2\eps}C_F
  \int \frac{d^dk}{(2\pi)^d}\,
  \frac{\gamma_\mu(\pthreeslash -\kslash +m)\Gamma
    (\ptwoslash -\kslash +m)\gamma^\mu}
       {k^2\big( (p_3-k)^2-m^2\big)\big( (p_2-k)^2-m^2\big)},\\
   \Lambda^{(1)}_{\Gamma,\xi\neq1} &= 
  ig^2\tilde\mu^{2\eps}C_F\,(1-\xi)
  \int \frac{d^dk}{(2\pi)^d}\,
  \frac{\kslash(\pthreeslash -\kslash +m)\Gamma
    (\ptwoslash -\kslash +m)\kslash/k^2}
       {k^2\big( (p_3-k)^2-m^2\big)\big( (p_2-k)^2-m^2\big)}.
\end{align}
\end{widetext}
The renormalization condition above requires us to compute
$\Tr[q\cdot\Lambda^{(1)}_\text{A}\gamma_5]$, with
$\Gamma^\nu_\text{A}=\gamma^\nu\gamma_5$. By tracing the numerators of
the two integrands for $\xi=1$ and $\xi\neq1$, we see that
\begin{equation}
  \Tr[\Lambda^{(1)}_\text{A,$\xi\neq1$}\gamma_5]
  = -\frac{1-\xi}d \Tr[\Lambda^{(1)}_\text{A,$\xi=1$}\gamma_5].
\end{equation}
Hence we have to evaluate only the $\xi=1$ (Feynman gauge) term for
$\Tr[\Lambda^{(1)}_\text{A,$\xi=1$}\gamma_5]$. Using the
notation,
\begin{equation}
  N_\Gamma = \gamma_\mu(\pthreeslash -\kslash +m)\,\Gamma\,
  (\ptwoslash -\kslash +m)\gamma^\mu,
\end{equation}
we have
\begin{equation}
    \Tr[q\cdot N_\text{A} \gamma_5] = 12 m d q^2
\end{equation}
and we learn that
$\Tr[q\cdot\Lambda^{(1)}_\text{A,$\xi=1$}\gamma_5]$ can be expressed
in terms of the finite integral
\begin{multline}
  \label{eq:usefulC0}
  -ig^2\tilde\mu^{2\eps}C_F\!\!
  \int\!\! \frac{d^dk}{(2\pi)^d}\,
  \frac1{k^2\big( (p_3{-}k)^2-m^2\big)\big( (p_2{-}k)^2-m^2\big)}\\
  =
  - \afpC \frac1{\mu^2} C_0(m^2/\mu^2),
\end{multline}
where $C_0(u)$ comes from a Feynman-parameter integral and is given by
\begin{widetext}

  \begin{equation}
  C_0(u) = \frac{2i}{\sqrt3} 
  \left[
    \text{Li}_2\Big(\frac{-i+\sqrt{3}}{\sqrt{3}-i \sqrt{4u+1}}\Big)
    \right.
    - \left.\text{Li}_2\Big(\frac{i+\sqrt{3}}{\sqrt{3}-i \sqrt{4u+1}}\Big)
    \right.
    + \left.
    \text{Li}_2\Big(\frac{-i+\sqrt{3}}{i\sqrt{4u+1}+\sqrt{3}}\Big)
    \right.
    - \left.\text{Li}_2\Big(\frac{i+\sqrt{3}}{i \sqrt{4u+1}+\sqrt{3}}\Big)
    \right].
\end{equation}
Since the result is finite, we can set $d=4$ and find
\begin{equation}
  \label{eq:TrqLamdaAg5}
  \Tr[q{\cdot}\Lambda^{(1)}_\text{A}\gamma_5]_\text{sym}
  = \Big(1-\frac{1-\xi}4\Big)
  \Tr[q{\cdot}\Lambda^{(1)}_{\text{A},\xi=1}\gamma_5]_\text{sym}
  = 12m \afpC (3+\xi) C_0(m^2/\mu^2).
\end{equation}
Now we have all we need to evaluate $\Zmmsmom$ from the
renormalization condition in~(\ref{eq:Z_m-minkowski}), which we
rewrite as
\begin{equation}
  \label{eq:Zm-mSMOM-cond}
  Z_m^\text{mSMOM} = \lim_{m_\text{R}\to\mbar}
  \frac1{12m}\frac1{Z_q^\text{mSMOM}}
  \left[
    \Tr\big({-i}S(p)^{-1}\big)_{p^2=-\mu^2} -
    \frac12 Z_\text{A}^\text{mSMOM}
    \Tr(q\cdot\Lambda_\text{A}\gamma_5)_\text{sym}
    \right].
\end{equation}
Using the results in~(\ref{eq:sigmap-xi}), (\ref{eq:Zqmsmom-xi})
and~(\ref{eq:TrqLamdaAg5}), together with (in Minkowski space) $i
S(p)^{-1} = \pslash - m - \Sigma(p)$, shows that to one loop,
\begin{equation}
  \begin{aligned}
  \Zmmsmom &= 1 + \afpC \left[ A_\xi + B_\xi
    - \frac{3+\xi}2 C_0(\mbar^2/\mu^2)\right]\\
      &= 1 + \afpC \left[
        3\frac1\epsbar +(4+\xi) - \frac{3+\xi}2\,C_0(u)
        - 3\ln(\mu^2/\tilde\mu^2)\right.\\
        &\phantom{= 1 + \afpC}
        \left.{}\; + \xi\bigg(u+u^2\ln\Big(\frac u{1+u}\Big)\bigg)
        +(3+\xi)\, u \ln\Big(\frac u{1+u}\Big) -3 \ln(1+u)
        \right],
\end{aligned}
\end{equation}
where now $u=\mbar^2/\mu^2$. Finally, the conversion factor is:
\begin{equation}
  \begin{aligned}
    R_m^{\msbar\leftarrow \text{mSMOM}} &= 1 + \afpC \left[
      -(4+\xi) + \frac{3+\xi}2\,C_0(u)
      + 3\ln(\mu^2/\tilde\mu^2)+ 3 \ln(1+u)\right.\\
      &\phantom{= 1 + \afpC}
      \left.{}\; -\xi\bigg(u+u^2\ln\Big(\frac u{1+u}\Big)\bigg)
      -(3+\xi)\, u \ln\Big(\frac u{1+u}\Big) 
      \right].
  \end{aligned}
  \label{eq:mSMOMtoMSbar}
\end{equation}
When $\bar m\to0$ ($u\to0$), this agrees with equation~(24) in Sturm
et al~\cite{Sturm:2009kb}, after setting $\tilde\mu=\mu$. For $\xi=1$
the result for $\ZPmsmom=1/\Zmmsmom$ reproduces the Feynman-gauge
result in Ref.~\cite{Boyle:2016wis}. We also computed $\ZPmsmom$
to one-loop order directly from the renormalization condition of equation~(\ref{eq:Z_P})
(in Minkowski space) and confirmed that $\ZPmsmom=1/\Zmmsmom$ for
arbitrary $\xi$.

The lattice computations are performed using Landau-gauge-fixed configurations and hence we need the conversion factor in Landau gauge, $\xi=0$:
\begin{equation}
  \label{eq:conv-factor-landau}
  R_m^{\msbar\leftarrow \text{mSMOM}} = 1 + \afpC \left[
    -4 + \frac32\,C_0(u) + 3\ln(\mu^2/\tilde\mu^2) + 3\ln(1+u) -3 u \ln\Big(\frac u{1+u}\Big) 
        \right].
\end{equation}
In figure~\ref{fig:mass-conv-factor}, we show plots of the mass conversion
factor as a function of $\mbar$, in both Feynman and Landau gauge for
two choices of matching scale $\mu$ (taking $\tilde\mu=\mu$).
\end{widetext}

\FloatBarrier
\bibliography{paper.bib}
\end{document}